\documentclass[twocolumn]{IEEEtran}

\usepackage[noadjust]{cite}
\ifCLASSINFOpdf
   \usepackage[pdftex]{graphicx}
  \DeclareGraphicsExtensions{.pdf,.jpeg,.png}
\else
  \usepackage[dvips]{graphicx}
   \DeclareGraphicsExtensions{.eps}
\fi
\usepackage[cmex10]{amsmath}
\usepackage{amssymb,amsfonts}

\begin{document}
\title{General Auction-Theoretic Strategies for Distributed Partner Selection in Cooperative Wireless Networks}

\author{Amitav~Mukherjee,~\IEEEmembership{Graduate~Student~Member,~IEEE,}
        and~Hyuck~M.~Kwon,~\IEEEmembership{Senior~Member,~IEEE}% <-this % stops a space
%\thanks{* Corresponding author}
\thanks{Amitav Mukherjee is with the Department of Electrical Engineering and Computer Science, University of California, Irvine, CA 92697 USA (e-mail: a.mukherjee@uci.edu)}
\thanks{Hyuck M. Kwon is with the Department of Electrical Engineering and Computer Science, Wichita State University, Wichita,
KS, 67260 USA (e-mail: hyuck.kwon@wichita.edu).}% <-this % stops a space
\thanks{This paper was presented in part at the 2008 IEEE Military Communications Conference (MILCOM), San Diego, CA USA.}% <-this % stops a space
\thanks{This work was partly sponsored by the Army Research Office under DEPSCoR ARO Grant W911NF-08-1-0256, and by NASA under EPSCoR CAN Grant NNX08AV84A.}
}

%\markboth{IEEE TRANSACTIONS ON COMMUNICATIONS,~Vol.~X, No.~X, XXXX~2010}%
%{Mukherjee \MakeLowercase{\textit{et al.}}: GENERAL AUCTION-THEORETIC STRATEGIES FOR DISTRIBUTED PARTNER SELECTION IN Cooperative WIRELESS NETWORKS}
\maketitle

%\IEEEpeerreviewmaketitle
\begin{abstract}
%Cooperative diversity is a form of distributed spatial diversity achieved by collaborative communications between mobile users in cellular or wireless data networks.
It is unrealistic to assume that all nodes in an ad hoc wireless network would be willing to participate in cooperative communication, especially if their desired Quality-of-Service (QoS) is achievable via direct transmission.
An incentive-based auction mechanism is presented to induce cooperative behavior in wireless networks with emphasis on users with asymmetrical channel fading conditions. A single-object second-price auction is studied for cooperative partner selection in single-carrier networks. In addition, a multiple-object bundled auction is analyzed for the selection of multiple simultaneous partners in a cooperative orthogonal frequency-division multiplexing (OFDM) setting. For both cases, we characterize equilibrium outage probability performance, seller revenue, and feedback bounds. The auction-based partner selection allows winning bidders to achieve their desired QoS while compensating the seller who assists them. At the local level sellers aim for revenue maximization, while connections are drawn to min-max fairness at the network level. The proposed strategies for partner selection in self-configuring cooperative wireless networks are shown to be robust under conditions of uncertainty in the number of users requesting cooperation, as well as minimal topology and channel link information available to individual users.
\end{abstract}
\section{Introduction}

\IEEEPARstart{T}{he} usage of multiple-input multiple-output (MIMO) systems for enhancing achievable data rates in wireless systems by means of spatial diversity has been a topic of interest over the last decade \cite{Foschini}. Spatial diversity with multiple antennas may be difficult to implement in space and power-constrained mobile terminals, which has led to the recent development of cooperative diversity \cite{Sendoris1}--\cite{Horst}. In brief, cooperative diversity is achieved when the data of a user is transmitted by its cooperating partners and vice-versa, which emulates a multi-antenna transmission scheme. This cooperation allows either an increase in achievable rates, a decrease in net power consumption, or an improved decoding performance due to diversity.

Conventional cooperation protocols are usually reciprocal, i.e., when cooperation between two users is mutually beneficial in terms of increased throughput or energy efficiency. For this purpose, amplify-and-forward (AF) cooperation involves the partner forwarding a scaled version of its received signal to the final destination; decode-and-forward (DF) cooperation requires the partner to decode and then retransmit its estimate of the received information \cite{Sendoris1}--\cite{Laneman}, and coded cooperation entails the alternate transmission of each other's incremental redundancy to a common destination \cite{Hunter}--\cite{Stefanov3}. Cooperative protocols are called full-duplex if partners transmit and receive information simultaneously over the same frequency, and half-duplex if partners are constrained to transmit and receive in channels that are orthogonal in time, frequency, or spreading code. While the AF and DF protocols offer different diversity and bit error benefits depending upon the quality of the inter-user channel, we adopt the full-duplex DF technique as the underlying relaying scheme since we assume that the inter-user channel is statistically better than the direct channels.

The selection of appropriate cooperative partners has been based largely on the optimization of physical layer parameters, primarily by a central controller. Examples include using a maximum received SNR criterion in \cite{Stefanov3}, a centralized utility maximization method in \cite{Zhang}, distributed selection using a fixed-priority list protocol in \cite{Hunter2}, maximum harmonic mean of source-relay and relay-destination channel gains in \cite{Liu08}, and transmit power-minimizing partner/subcarrier assignment in \cite{Shen08,Liu09}. However, the majority of partner-selection research has been relay-oriented in nature, meaning the entire network is searched for optimal partners, and it is assumed that every user is involved in cooperative transmission over the network lifetime. Another common assumption is that data sources have ready access to fixed relays in the network that do not have any information of their own to transmit.

In practice, even in small networks the rapidly changing inter-user channels, user asynchronicity, diverse Quality-of-Service (QoS) requirements, user mobility, and users entering and leaving randomly make it difficult for users to gather enough information to even decide if cooperation is beneficial. More importantly, self-interested users may not always want to voluntarily engage in cooperation, particularly with weaker users in asymmetric fading conditions. Therefore, in this paper we make the more realistic assumption that users seek relaying assistance only if they are unable to achieve a desired QoS on their own. %We then construct a user-centric partner-selection approach with incentives to cooperate, where local users individually decide the need and duration of cooperative transmissions.
%It must be noted that we do not consider pure relay channels or altruistic cooperation, since all cooperating users have their own data to transmit.
We then develop an auction-theoretic cooperative partner selection (ACOPS) scheme by modeling users seeking cooperation as bidders and potential helpers as sellers. The use of auctions provides an incentive for stronger users to share their resources based on monetary incentives.
%We seek to address one of the open research problems mentioned in \cite{Sendoris1}, the pioneering paper on cooperative communications.
Previous applications of microeconomics to wireless networks include the use of pricing functions for regulation of power control, packet forwarding, and spectrum allocation in cognitive networks \cite{Mandayam1}--\cite{Hossain08}.

The characteristics of competition and collaboration in cooperative networks naturally invite analyses based on game theory \cite{Kishore08,Sankar08}. Auction-theoretic algorithms (a subset of game theory) have been proposed in \cite{Maille} for centralized downlink power allocation in code-division multiple-access systems, in \cite{JunSun} for downlink bandwidth allocation in wireless fading channels, in \cite{Poor} for iterative power allocation by relays operating with a half-duplex AF protocol, and \cite{Han} for bandwidth allocation on an OFDMA downlink. This paper focuses on the decentralized cooperative partner selection on the uplink of cellular or ad hoc networks using a second-price auction structure, as compared to the downlink resource allocation in the previous work mentioned above. The proposed auction approach is an example of so-called \emph{competitive} fairness, since the users compete to decide the recipient of cooperative resources. In comparison, the application of pricing functions as in \cite{Mandayam1}--\cite{Mandayam2} imposes a central utility function upon all users, which has been criticized as artificial fairness \cite{JunSun}.

The contributions of this paper are the following:
\begin{itemize}
\item We characterize best-reply strategies and seller revenue for the single-stage second-price sealed-bid auction in which the exact number of competitors is unknown for each bidder, and a single user is assisted over one network realization. We then prove the existence and uniqueness of the symmetric Nash equilibria for the auction game under consideration. Finally, we calculate the probability of outage and signaling overhead bounds in the context of single-partner cooperative communications over flat-fading channels.
\item We show that selling all objects together in a one-time multiple-object auction \cite{Krishna} for multiple cooperative partner selection is preferable to holding sequential single-object auctions. We characterize revenue, probability of outage, and signaling overhead for a bundled multiple-object auction. This selection scheme is examined in the context of cooperative orthogonal frequency division multiplexing (OFDM) communications \cite{Tarokh}--\cite{LiuOFDM08} over frequency-selective fading channels.
\item Finally, we develop a complete framework for the application of the novel results obtained above to a cooperative communication network. The performance of the proposed algorithms is then compared with existing partner selection, fully centralized partner allocation, and no cooperation schemes.
\end{itemize}

The remainder of this paper is organized as follows. Section~\ref{sec:SysModel} describes the cooperative communication system framework. Section~\ref{sec:AuctionTheory} describes the auction-theoretic model used in subsequent sections. Section~\ref{sec:SinglePartner} introduces the single-stage second-price auction for the selection of a single partner per cooperative transmission. Section~\ref{sec:MultiPartner} presents a multiple-object bundle auction for multiple-partner selection in multicarrier systems, followed by a discussion on extensions to sequential auction games and multi-hop relaying in Section~\ref{sec:Extend}. Finally, simulation results are presented in Section~\ref{sec:Simul}, and Section~\ref{sec:Conclus} concludes the paper.

\emph{Notation}: Random variables are represented in italics, e.g., \emph{Y}, while their realizations are in lower-case, \emph{y}. Vectors are in boldface, ${\bf{v}}$. Sets are denoted by mathematical script, ${\cal A}$. The Cartesian product is denoted by $\times$, convolution by $*$, the expectation operation by $E\left[  \cdot  \right]$, and the Laplace transform by ${\mathcal{L}} \left\{  \cdot  \right\}$.

\section{COOPERATIVE SYSTEM MODEL}\label{sec:SysModel}
The wireless network under consideration consists of single-antenna users self-grouped according to geographical proximity. User antennas are assumed to be omni-directional to facilitate cooperative communications. Therefore, a group for a particular user may be approximated as a circle with radius limited by the range up to which signals can be transmitted and received with neighboring users, unlike \cite{Hunter2} where users must be able to receive signals from all other users.

We focus on a cellular network with a single destination, and assume that the users may be grouped into two categories based upon their uplink channel condition. Users that are unable to achieve a desired transmission rate due to poor channel quality to the destination or BS are termed \emph{weak users}, and users possessing favorable channels to the BS are referred to as \emph{potential helpers} or sellers in the remainder of this paper. In principle, the proposed scheme is applicable to a broad class of networks, for example, ad hoc wireless networks with differing source-destination pairs as compared to a fixed BS in cellular systems.

The proposed ACOPS model makes the following major assumptions:
\begin{enumerate}
 \item[(\emph{A}1)] Each user is assumed to have knowledge only of the inter-user link CSI between itself and the potential helper within its group, and the CSI of its direct link to the BS,
 \item[(\emph{A}2)] Weak users and helpers are both unaware of the exact number of neighboring weak users and their channel conditions,
 \item[(\emph{A}3)] A single potential helper (seller) exists per group,
 \item[(\emph{A}4)] Only a single (one-shot) cooperation interval is analyzed; and all cooperative communications are carried out over a single hop.
\end{enumerate}
Assumptions (\emph{A}1)-(\emph{A}2) highlight the lack of complete information in the proposed partner-selection scheme, while (\emph{A}3)-(\emph{A}4) enable a simpler starting point for our investigation. Extensions to (\emph{A}3)-(\emph{A}4) such as multiple potential helpers per group and cooperation over multiple stages are discussed in Section~\ref{sec:Extend}. In particular, Sec.~\ref{sec:Extend} describes the considerable challenges that are introduced when examining a multi-stage version of the auction game in realistic wireless networks with dynamic user channels.

\subsection{Cooperation over Flat-Fading Channels}
Let ${\cal N} = \left\{ {1,2, \ldots ,N} \right\}$ be the set of all weak users in a single group. All direct-to-BS links and inter-user links between potential helpers and weak users within a group are modeled using a combination of small-scale fading, path loss, and shadowing \cite[sec. 3.2]{Goldsmith}. The small-scale fading is quasi-static Rayleigh fading in nature. For the single-antenna user case, the channel capacity for a link with SNR $\gamma$ and a Gaussian input is
\begin{equation}
C\left( \gamma  \right) = \log _2 \left( {1 + \gamma } \right){\text{ bits/s/Hz}},
\label{EQ:Capacity}
\end{equation}
subject to an average transmit power constraint $P_T$. The instantaneous SNR of the link between users \emph{i} and \emph{j} at a distance  $d_{i,j}$ can be represented as $\gamma _{i,j}  = \Gamma _{i,j} \left| {h_{i,j} } \right|^2$, where $h_{i,j}$ is the normalized complex Gaussian fading coefficient with amplitude $\left| {h_{i,j} } \right|^2  = \left( {h_{i,j,c} } \right)^2  + \left( {h_{i,j,s} } \right)^2$, $h_{i,j,c} ,h_{i,j,s}  \sim \mathcal{N}\left( {0,{1 \mathord{\left/ {\vphantom {1 2}} \right. \kern-\nulldelimiterspace} 2}} \right)$ being normally distributed, and $E\left\{ {\left| {h_{i,j} } \right|^2 } \right\} = 1$. $\Gamma _{i,j}  = \frac{{P_T }}{{N_0 W}}G_T G_R s_{i,j} d_{i,j}^{ - a }$ is the average SNR of the link, with transmit power $P_T$, bandwidth \emph{W}, transmit and receive antenna gains $G_T, G_R$, log-normally distributed shadowing component $s_{i,j}$, and path-loss exponent $a$, $2 \le a  \le 4$ typically. Complex additive white Gaussian noise with zero mean and variance ${{N_0 } \mathord{\left/ {\vphantom {{N_0 } 2}} \right. \kern-\nulldelimiterspace} 2}$ per dimension is present at all receiving terminals. The direct-to-BS links of the weak users are assumed to be dominated by the large-scale fading effects of shadowing and path loss, over an extended period of time.

Let $R_{i,BS}$ denote the achievable direct-link transmission rate without cooperation for the $i^{th}$ weak user. Assume the desired transmission rate to be $D_i$, $D_i  > R_{i,BS}$, which indicates a shortfall of $q_i  = D_i  - R_{i,BS}$. User \emph{i} then seeks to approach the target rate $D_i$ by attempting to cooperate with a potential helper in its vicinity, if any. Under Rayleigh fading, $\gamma _{i,j}$ is exponentially distributed with parameter ${1 \mathord{\left/ {\vphantom {1 {\Gamma _{i,j} }}} \right. \kern-\nulldelimiterspace} {\Gamma _{i,j} }}$. Therefore, if we define the outage probability for a target rate $D_i$ as the probability of $D_i$ exceeding the achievable link capacity $C\left( \gamma  \right)$, i.e., $P_{out} \left( {D_i } \right) = \Pr \left( {C\left( \gamma  \right) < D_i } \right)$, then from (\ref{EQ:Capacity}) we can write the non-cooperative outage probability as
\begin{eqnarray}
P_{out} \left( {D_i } \right) &=& \int\limits_0^{2^{D_i }  - 1} {\frac{1}{{\Gamma _{i,j} }}} \exp \left( { - \frac{\gamma }{{\Gamma _{i,j} }}} \right)d\gamma \nonumber\\
  &=& 1 - \exp \left( { - \frac{{2^{D_i }  - 1}}{{\Gamma _{i,j} }}} \right).
\label{EQ:NonCoopOutage}
\end{eqnarray}
The achievable total rate $R_i$ for the $i^{th}$ weak user cooperating with a single potential helper is \cite{Sendoris1}
\begin{equation}
R_i  = R_{i,PH}  + R_{i,BS}, \label{EQ:Cooprate}
\end{equation}
where $R_{i,PH}$ denotes the cooperative rate provided by the potential helper to the $i^{th}$ weak user. For simulation purposes we assume an idealized full-duplex decode-and-forward protocol with perfect echo cancelation and synchronization, as well as zero processing delay. If half-duplex cooperation is considered, a suitable factor of $\frac{1}{2}$ can be introduced in (\ref{EQ:Cooprate}) to account for the rate penalty incurred. The specific cooperative rate expressions for AF and DF protocols with a half-duplex relaying constraint have been derived in \cite{Laneman}, and can be integrated into ACOPS if so desired.

A weak user \emph{i} that desires a higher uplink data rate broadcasts its cooperative rate request $q_i$, which is overheard by neighboring users.
%For simplicity, it is assumed that there are no collisions between the rate requests of co-located weak users.
Potential helpers who receive such request(s) respond with the data rate $R_c$ they are willing to allocate for cooperative communication, $R_c  \leqslant C\left( {\gamma _{PH,BS} } \right)$. The helper-user links are considered to be reciprocal; this response allows the weak users to estimate the quality of their link to the helper, fulfilling Assumption (\emph{A}1).

\subsection{Cooperation over Frequency-Selective Fading Channels}
For a cooperative OFDM scenario, the wireless link between users is modeled as a frequency-selective block-fading channel with $L$ delay paths or taps and an exponentially decaying intensity profile. If the total bandwidth is divided into $\tilde K$ subchannels with uniform power allocation to decrease CSI requirements, the total capacity is the sum of the rates of the parallel subchannels as
\begin{equation}
\tilde C_{i,j} \left( {\tilde K} \right) = \sum\limits_{k = 0}^{\tilde K - 1} {C\left( {\gamma _{i,j}^k } \right) = } \sum\limits_{k = 0}^{\tilde K - 1} {\log _2 } \left( {1 + \gamma _{i,j}^k } \right), \label{EQ:OFDMCapacity}
\end{equation}
where $\gamma _{i,j}^k$ is the exponentially distributed SNR on subchannel \emph{k} between users $i$ and $j$ and implicitly depends on $L$. Our cooperative OFDM model considers a single potential helper with $\tilde K$ subchannels offered for cooperation to \emph{N} weak users, $N \le \tilde K$ since subchannels are not allowed to be shared between users over one cooperative epoch.

\begin{figure}[ht]
\centering
\includegraphics[width=\linewidth]{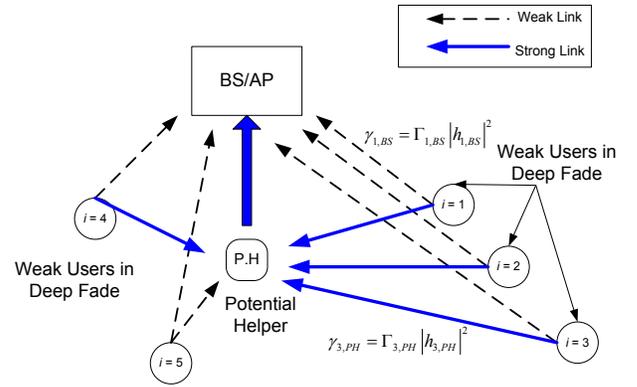}
\caption{Cooperative wireless network with asymmetric link conditions for a group with a single potential helper and $N = 5$ weak users.}
\label{fig:Cell}
\end{figure}
An example of an asymmetric single-carrier cooperative network is shown in Fig. 1, where a single potential helper and \emph{N} = 5 weak users are assumed to constitute a group. All five weak users are assumed to be in outage and therefore broadcast cooperative rate requests $q_1 , \ldots ,q_5$. Four of the five weak users, numbered as \emph{i} = 1, 2, 3, and 4, possess a much more favorable link to the potential helper compared to their direct links to the BS, and therefore receive a response from the potential helper which has a strong link to the BS, i.e., $E\left| {h_{PH,BS} } \right| \gg E\left| {h_{i,BS} } \right|$. The feedback from the potential helper enables the weak users to estimate the quality of their user-to-helper link $\gamma _{i,PH}$. The fifth weak user (\emph{i} = 5), in Fig. 1 has unfavorable channels to both the BS and the potential helper. Therefore it may not receive a response from the potential helper at all, or based upon the response, it perceives the poor quality of its user-to-helper link and abstains from cooperation.

\section{AUCTION-THEORETIC BACKGROUND}\label{sec:AuctionTheory}

\subsection{Auction as a Game}
Auctions can be modeled as games of incomplete information \cite{Krishna},\cite{Myerson}. A game $\Omega$ consists of (i) a set of $\cal N$  players, (ii) for each player $i \in \cal N$, a nonempty set of actions $\mathcal{S}_{i}$ and a payoff function $\Pi_i$. Therefore, an auction game with player set $\mathcal {N}$ can be compactly represented in strategic form as $\Omega = \left[{\mathcal{N}, \mathcal{S}_{i}, \Pi_{i}}\right]_{i \in {\mathcal{N}}}$.

In a sealed-bid auction, bids are submitted confidentially and are known only to the seller. A first-price sealed-bid auction awards the object to the highest bidder who pays his bid, whereas a second-price auction awards the object to the highest bidder who pays an amount equal to the second-highest of all bids placed. In both formats, losing bidders do not pay anything. It is shown in subsequent sections that the proposed auction game strikes a balance between revenue maximization at the seller level, and efficient allocation at the network level.

%We restrict our attention to the revenue criteria since the potential helper's local perspective is revenue-oriented, even though network-wide efficiency may be more desirable from a centralized viewpoint. %We note that non-standard auctions, such as the all-pay auction, may offer even higher expected seller revenue in some very specific situations \cite{Krishna}, but we focus on the well-known standard auctions, such as sealed-bid first-price and second-price auctions without reserve prices, for ease of exposition.

All auction-theoretic mechanisms in this paper conform to standard auctions in the following ways: (1) the object(s) on sale is always awarded to the highest bidder, (2) all bidders are treated as anonymous entities, and (3) ties are broken by random allotment of the object to one of the high bidders. Our assumptions for ACOPS, such as user asynchronicity, uncertainty about the number of bidders and private values signify that we analyze auctions as games of incomplete information. The relevance of objects, payoffs, bids, and winners for the cooperative wireless network under consideration is explained later in this section. Even though bids are broadcast and can be overheard by competing bidders, the assumption of private values still holds since bids already placed cannot be updated within one contention interval, and it is shown in the next section that the optimum bidding strategy is independent of other bids. In other words, bids that have not yet been placed are not influenced by overheard bids, and hence the auction is equivalent to a sealed-bid format.

At user initiation, each user is provided with an equal amount of virtual `money' or tokens for the purpose of resource contention.%which have no physical significance other than the purpose of cooperative contention.
The proposed wireless network model has a group of weak users willing to pay varying amounts of `money' for higher degrees of Quality of Service (QoS), and a group of potential helpers willing to receive monetary compensation for sharing their resources and assisting such users as envisioned in \cite{Sendoris1}. Revenue accumulated by helpers can be used in turn for their own cooperative bids during adverse link conditions in the future, since as stated before, potential helpers are not fixed over time. It is in the best interest of every potential helper to maximize its expected revenue. Therefore, the proposed mechanism serves as an incentive for advantaged users to participate in cooperation over the network.

\subsection{Contention Process}
\begin{figure}[ht]
\centering
\includegraphics[width=\linewidth]{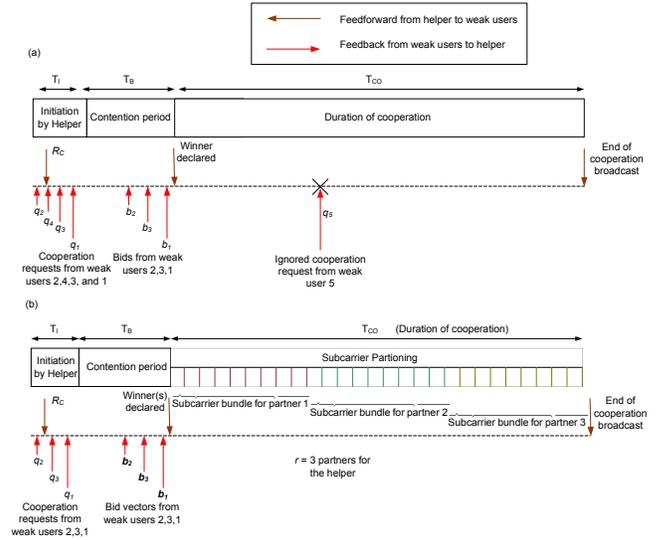}
\caption{(a) Cooperative communication control structure for single-partner selection using a single-stage auction with $N = 4$ weak users and $N_a = 3$ bidders involved in the contention stage. Cooperation requests $q_i$ and bids $b_i$ are received at different times by the helper due to asynchronous users. The duration of cooperation denoted by $T_{CO}$ is limited by the channel coherence time of the potential helper; since the helper itself is mobile it may not always be in a more advantageous position with respect to the BS. Both $T_I$ and $T_B$ are assumed to be fractions of $T_{CO}$, i.e., $T_I  \approx T_B  <  < {\text{ }}T_{CO}$. (b) Cooperative OFDM communication control structure for multiple-partner selection using a mixed-bundle multiple-object auction with $N_a = 3$ bidders involved in the contention stage.}
\label{fig:BidTime}
\end{figure}

Fig. 2 depicts a time-scale showing the initiation of the auction protocol during time $T_I$, where the potential helper is contacted by \emph{N} weak users belonging to a group, all of whom seek to benefit from cooperation with it. Based upon the requested data rates $q_i$ and its own surplus data rate $R_c$ ($R_c$ being the object on sale), the potential helper calculates the number of users it can assist simultaneously. The auction contention occurs if and only if (a) the number of cooperation requests received by the potential helper is greater than or equal to two, and (b) if $R_c > 0$, i.e., the helper's own QoS is already satisfied. The contention process takes place during interval $T_B$, followed by cooperative communication between the auction winner and the helper for time period $T_{CO}$. Once the bidding stage with a \emph{hard close time} $T_B$ is over and cooperative transmissions have commenced, the  helper ignores any cooperative requests received during time $T_{CO}$. Since the users are asynchronous, therefore cooperation requests and bids are received at different times by the potential helper within the initiation and contention periods, as seen in Fig. 2.

\subsection{Bidder Private Information}
\begin{table}[htbp]
\caption{Summary of Parameters}
\label{table:Params}
\centering
%\newpage
\begin{tabular}{|c|c|}
\hline
$X_i$ & Private value random variable for $i^{th}$ user\\
\hline
$N$ & Total number of weak users in a group\\
\hline
$N_a$ &	Number of actual bidders in a group\\
\hline
$R_c$ &	Potential helper surplus rate for sale\\
\hline
$q_i$ &	$i^{th}$ weak user rate request\\
\hline
% & Estimate of number of bidders by ith user
$b_i$ &	Bid placed by bidder \emph{i}\\
\hline
$r$ &	Number of simultaneous partners\\
\hline
$Y^k_i$ & $k^{th}$ bundle value for user \emph{i}\\
\hline
$c_k$ & Cardinality of bundle \emph{k}\\
\hline
\end{tabular}
\end{table}

Table~\ref{table:Params} lists the commonly used variables in this paper. Let random variable $\emph{X}_{i}$ represent the value of the sale object to user \textit{i}, where each $\emph{X}_{i}$ is independently and identically distributed with a continuous distribution function (cdf) $\emph{F}_X$ over support\footnote{Although negative private values seem unreasonable, we connect our private value to a physical parameter, namely the difference of two SNRs as explained in the next section.} $\left( {-\infty,\infty } \right)$. The private values $\emph{X}_{i}$ are considered to be independent, and are therefore unaffected by knowledge of values held by the other users.

Let $N_a$ be the number of \emph{actual} bidders during contention period $T_{B}$, where $N_a  \leq N$. The number of actual bidders may be less than the total number of weak users ${N}$, due to weak users with non-positive private values opting to not bid, improvements in a weak user's BS link quality $\gamma _{i,BS}$ after their cooperation request, or weak users moving out of range into another group. More precisely, a weak user proceeds to act as a bidder if it possesses a positive private value $\emph{x}_{i} > 0$. The entry of new weak users in the group during period $T_{B}$ will not change set $\cal N$ as they lack the information required to place a bid. Bidder \emph{i} estimates the number of competitors to be $\hat N_{a\left( i \right)}  \in \mathbb{Z}_+$, based upon the number of cooperation requests received by him/her. This estimate $\hat N_{a\left( i \right)}$ is modeled as a Poisson random variable with mean $\zeta  = E\left[ {N_a } \right] = E\left[ {\hat N_a } \right]$  and standard deviation $\sigma  = \sqrt \zeta$, i.e., $\Pr \left( {\hat N_a  = a} \right) = {{\zeta ^a e^{ - \zeta } } \mathord{\left/ {\vphantom {{\zeta ^a e^{ - \zeta } } {a!}}} \right. \kern-\nulldelimiterspace} {a!}}$, where $a \in \mathbb{Z}_ +$. Therefore, the seller and all bidders are uncertain about the ex ante (i.e., \emph{a priori}) number of actual bidders during the contention period $T_I$, in contrast to the existing auction-theoretic applications, such as \cite{JunSun}--\cite{Poor}. We refer to user uncertainty about the exact number of bidders as \emph{rival uncertainty}. We then define the (private value, rival estimate) 2-tuple of bidder $i$ as type $\left( {x_i ,\hat N_{a\left( i \right)} } \right)$ for a particular network realization, with $x_i  \in \mathbb{R}$.

Let an Euclidean subspace $\mathcal{S}_{i}$ be the strategy set of continuous, real-valued monotonically increasing bidding functions over the support of $\emph{X}_{i}$ for the $i^{th}$ bidder, % as ${\mathcal{S}}_i  = \{ b_i^{(1)} ,...,b_i^{(n)} \} _{i \in {\mathcal{N}}}$
with higher bidding functions corresponding to more favorable perceived private values. The joint strategy space for all \emph{N} users is then defined by the Cartesian product ${\mathcal{S}} = {\mathcal{S}}_1  \times {\mathcal{S}}_2  \times  \ldots  \times {\mathcal{S}}_N$. In addition, let vector   ${\mathbf{b}}_{ - i}  = \left\{ {b_1  \in {\mathcal{S}}_1 , \ldots ,b_{i - 1}  \in {\mathcal{S}}_{i - 1} ,b_{i + 1}  \in {\mathcal{S}}_{i + 1} , \ldots ,b_N  \in {\mathcal{S}}_N } \right\}$ denote bidding strategies for all \emph{N}-1 other users excluding bidder \emph{i}. The reserve price of the potential helper is set to zero for every object, i.e., the surplus data rate $R_c$ holds no value for the seller. Define the constant $\alpha_i   \triangleq \frac{{q_i }}{{R_c }}$, such that $\frac{1}{\alpha_i}$ denotes the ratio of the offered cooperative rate to the desired additional data rate of each user.

For any multi-user game, the payoff or profit function $\Pi _i$ for user \emph{i} represents the expected benefit or utility given a specific strategy (bid in the case of an auction game). Every bidder seeks to maximize its individual expected payoff, which is termed as risk-neutral behavior in game-theory literature.

\section{SINGLE-PARTNER SELECTION}\label{sec:SinglePartner}
In this section, we assume that the surplus rate $R_c$ of a potential helper is sufficient to support a single weak user, i.e., a single partner, at a given time. This situation corresponds to single-carrier AF/DF cooperation over a flat-fading channel, or a coded cooperative system with two users. First, we examine a single-stage game employing a second-price auction with \emph{N} weak users each facing an uncertain number of competitors.

The formulation of the private value for a bidder is an interesting design problem in itself. Consider the following two options out of the many possibilities: (a) $\emph{X}_i  = \gamma _{i,PH}$, or (b) $X_i  = \gamma _{i,PH}/\alpha_i  - \gamma _{i,BS}$. Here, (a) would be equivalent to a conventional partner selection strategy based upon maximizing the instantaneous helper-user SNR as in \cite{Zhang}, which is a greedy or \emph{opportunistic} approach.
On the other hand, the interpretation of (b) is more subtle. Theoretically, (b) is a measure of bidder \emph{i}'s marginal opportunity cost of not being awarded cooperation \cite{Nicholson}. The factor of $\frac{1}{\alpha_i}$ represents the utility perceived by the bidder; if $R_c < q_i$, then the bidder intuitively places a lower value on the sale object, and vice versa. Qualitatively speaking, this formulation of the private value also captures different measures of fairness depending on the state of the network. Consider a network with two weak users $A$ and $B$ with the same desired rate, and one helper. We now illustrate the tradeoff in fairness with two simple examples, assuming $\alpha_i=1 \quad\forall i$.

\emph{Example 1}: $A$ and $B$ have an equally poor direct link to the BS such that $\gamma_{A,BS}=\gamma_{B,BS}$. If $A$ has a better link to the helper, then $A$ wins the auction by virtue of having the highest bid. This outcome is equivalent to an opportunistic scheduling that would also take place if (a) was set as the private value.

\emph{Example 2}: $A$ and $B$ have an equally strong link to the helper such that $\gamma_{A,PH}=\gamma_{B,PH}$. If $B$ has a weaker link to the BS, then $B$ wins the auction due to a greater private value. This case now corresponds to a \emph{max-min} fair scheduling, since if $A$ was awarded cooperation, the minimum system rate $R_B=R_{B,BS}$ achieved by the loser would be lower (as $R_{B,BS}<R_{A,BS}$).

In a larger network with more weak users and multiple helpers interacting over multiple bidding intervals, the net effect of the private value based on (b) would tend towards a max-min fair outcome, since the weakest users have a greater chance of winning. Note that although users with very strong channels to the helper and a good channel to the BS will also have a large private value, the probability of their being in outage and participating in the auction game is low. While opportunistic, max-min, and proportionally fair scheduling algorithms have been widely discussed for non-cooperative centralized wireless networks, fairness measures in cooperative and ad hoc networks have largely focused on energy efficiency \cite{Karagiannidis08} or coalitional game theory \cite{Han08}.

Furthermore, $\gamma _{i,PH}$ and $\gamma _{i,BS}$ are i.i.d exponential random variables with rate parameters ${1 \mathord{\left/ {\vphantom {1 {\Gamma _{i,PH} }}} \right. \kern-\nulldelimiterspace} {\Gamma _{i,PH} }}$ and ${1 \mathord{\left/ {\vphantom {1 {\Gamma _{i,BS} }}} \right. \kern-\nulldelimiterspace} {\Gamma _{i,BS} }}$, respectively. Therefore, $V_i  = \frac{{\gamma _{i,PH} }}{{\alpha _i }}$ has the probability density function (pdf) $f_{V_i } \left( {v_i } \right) = \alpha_i f_{\gamma _{i,PH} } \left( {\alpha _i v_i } \right)$.

Let $\lambda_i={{\Gamma _{i,PH}  + \Gamma _{i,BS} }}$.
 Utilizing the fact that the density of the sum of two independent random variables is equal to the convolution of their densities, we obtain the private value pdf $f_{X_i } \left( {x_i } \right)$ as
\begin{eqnarray}
f_{X_i } \left( {x_i } \right) &=& \alpha_i f_{\gamma _{i,PH} } \left( {\alpha _i x_i } \right) * f_{\gamma _{i,BS} } \left( {-x_i } \right)\\
&=& \frac{\alpha _i}{\lambda_i}
\left\{ {\begin{array}{*{20}c}
   {e^{ - \frac{1}
{{\Gamma _{i,PH} }}\alpha _ix_i } } & {x_i  > 0}  \\
   {e^{\frac{1}
{{\Gamma _{i,BS} }}\alpha _ix_i } } & {x_i  < 0}.  \\
 \end{array} } \right.
 \label{EQ:singlepdf}
\end{eqnarray}

The pdf in (\ref{EQ:singlepdf}) simplifies to the Laplace distribution for the special case ${1 \mathord{\left/ {\vphantom {1 {\Gamma _{i,PH} }}} \right.
 \kern-\nulldelimiterspace} {\Gamma _{i,PH} }} = {1 \mathord{\left/ {\vphantom {1 {\Gamma _{i,BS} }}} \right. \kern-\nulldelimiterspace} {\Gamma _{i,BS} }}$. The corresponding private value cumulative distribution function (cdf) $F_{X_i } \left( {x_i } \right)$ is given by
%\begin{equation}
%F_{X_i } \left( {x_i } \right)
%%=\int\limits_{ - \infty }^{x_i } {f_{X_i } \left( u \right)du }
% = \frac{1}
%{{\Gamma _{i,PH}  + \Gamma _{i,BS} }}\left\{ {\begin{array}{*{20}c}
%   {\left( {\Gamma _{i,PH}  + \Gamma _{i,BS} } \right) - \Gamma _{i,PH}  \cdot e^{ - \frac{1}
%{{\Gamma _{i,PH} }}\alpha_i x_i } } & {x_i  > 0}  \\
%   {\Gamma _{i,BS}  \cdot e^{\frac{1}
%{{\Gamma _{i,BS} }}\alpha_i x_i } } & {x_i  < 0}.  \\
% \end{array} } \right.
%\label{EQ:singlecdf}
%\end{equation}
\begin{equation}
F_{X_i } \left( {x_i } \right)
%=\int\limits_{ - \infty }^{x_i } {f_{X_i } \left( u \right)du }
 = \frac{1}{\lambda_i} \left\{ {\begin{array}{*{20}c}
   { \lambda_i - \Gamma _{i,PH}  \cdot e^{ - \frac{1}
{{\Gamma _{i,PH} }}\alpha_i x_i } } & {x_i  > 0}  \\
   {\Gamma _{i,BS}  \cdot e^{\frac{1}
{{\Gamma _{i,BS} }}\alpha_i x_i } } & {x_i  < 0}.  \\
 \end{array} } \right.
\label{EQ:singlecdf}
\end{equation}

We can then define the payoff for the $i^{th}$ user under second-price rules as
\begin{equation}
\Pi _i  \triangleq \left\{ {\begin{array}{*{20}c}
   {\underbrace {\gamma_{i,PH}/\alpha_i  - \gamma _{i,BS} }_{value} - \underbrace {\mathop {\max }\limits_{j \ne i} b_j }_{payment}} & {\text{if}\hspace{0.1in}b_i>\mathop {\max }\limits_{j \ne i} b_j },  \\
   0 & \hspace{0.1in}{\text{otherwise}},   \\
 \end{array} } \right. \label{EQ:payoff}
\end{equation}
where the monetary unit of payoffs, values, and bids is in terms of SNR.
\subsection{Equilibrium Strategy and Revenue}
We first define the pure-strategy best-response correspondence for bidder \emph{i} in the single-stage second-price auction game as $BR_i^s  = \arg \mathop {\max }\limits_{b_i  \in S_i } \Pi _i \left( {b_i ,{\mathbf{b}}_{ - i} } \right)$, $\forall {\text{ }}\tilde B = \left\{ {b_i ,{\mathbf{b}}_{ - i} } \right\} \in \mathcal{S}$, $\forall {\text{ }}i \in {\mathcal{N}}$, where the superscript $^{\emph{s}}$ indicates the single-stage auction game, and the subscript \emph{i} denotes bidder index. The symmetric equilibrium strategy is a joint strategy ${\mathcal{B}} = \left( {BR_1^s , \ldots ,BR_N^s } \right)$ such that no player can increase their payoff by unilaterally deviating to another strategy, and it is symmetric, all players adopt the same strategy.

We state some general characteristics of best-response correspondence strategies $BR_i \left( {x_i ,\hat N_{a\left( i \right)} } \right)$ in Proposition 1, with the superscript $^{\emph{s}}$ suppressed for convenience.

\emph{Proposition 1}: A best-response strategy in equilibrium must have the following characteristics: (a) $BR_i \left( {x_i  \leqslant 0,\hat N_{a\left( i \right)} } \right) = 0$; (b) $BR_i \left( {x_i ,\hat N_{a\left( i \right)} } \right)$ is non-decreasing with respect to $x_i$,
%i.e., $BR_i \left( {x_i^{\prime} ,\hat N_{a\left( i \right)} } \right) \geqslant BR_i \left( {x_i ,\hat N_{a\left( i \right)} } \right)$  for $x_i ^\prime   > x_i$;
and (c) the best reply correspondence $BR_i^{}$ for a user of type $\left( {X_i  = x_i ,\hat N_{a\left( i \right)}  = N_a } \right)$ is $BR_i  = b_i \left( {x_i } \right) = x_i$.

\emph{Proof}: In brief, (a) follows from boundary conditions to ensure that payoffs remain non-negative. (b) ensures that bidders with higher types bid higher and have a greater probability of winning. For (c), standard arguments as in \cite[Section 3.2.2]{Krishna} show that it is irrational to either over-bid or under-bid with respect to one's private value regardless of rival uncertainty. $\blacksquare$

Equilibrium strategies are of interest in any multi-user game since the existence of Nash equilibria (NE) is usually a desirable outcome. The common method of proving the existence of NE is by invoking the classical Kakutani or Debreu-Fan-Glicksburg fixed-point theorems \cite{Myerson},\cite{Tirole}. However, in order to apply these well-known methods, the user payoff functions must be continuous, as in \cite{JunSun}, where the user payoff was equated to the channel realization and therefore the Kakutani theorem was applicable. However, in the proposed auction game user payoffs are clearly discontinuous, as seen from (\ref{EQ:payoff}). Therefore, we invoke the Dasgupta-Maskin existence theorem for NE in games with discontinuous payoffs \cite{Dasgupta}.

\emph{Theorem 1}: A pure-strategy Nash Equilibrium exists for the proposed \emph{N}-bidder second-price auction game $\Omega = [{\mathcal{N}, \mathcal{S}_{i}, \Pi_{i}}]_{i \in {\mathcal{N}}}$ under rival uncertainty with the best-response correspondence function given in Proposition 1.

\emph{Proof}: From Proposition 1, since the best-response bidding strategy is to bid the private value itself, each Euclidean strategy subspace of valid bids $\mathcal{S}_{i}$ is a non-empty and convex set, and compact since it is closed and bounded. Therefore the overall strategy space $\mathcal{S}$ is the convex product of convex sets. The user payoff function $\Pi _i$ in (\ref{EQ:payoff}) is concave (i.e., convex cap) since it is a sum of two concave functions (the first term is $b_i$ which is concave in itself, and the second term is concave since the maximizer function is convex), which in turn implies quasi-concavity. $\Pi _i$ is upper semi-continuous, which can be shown by contradiction, as in \cite[pg. 38, Lemma 5]{Mackenzie}. Define $\Pi _i^* \left( {{\mathbf{b}}_{ - i} } \right) \equiv \mathop {\max }\limits_{b_i } \Pi _i \left( {b_i ,{\mathbf{b}}_{ - i} } \right)$. For a given $b_i$, $\Pi _i^* \left( {{\mathbf{b}}_{ - i} } \right)$ is a constant maximum value when it is non-zero and is continuous over ${\mathbf{b}}_i$, thus, $\Pi _i$ has a continuous maximum. Therefore, the Dasgupta-Maskin conditions are satisfied, and the existence of a pure-strategy Nash Equilibrium is guaranteed. $\blacksquare$

\emph{Proposition 2}: The Nash equilibrium described above is the unique, symmetric, and efficient equilibrium of the proposed ACOPS game.

\emph{Proof}: As described earlier, the ACOPS game conforms to a standard sealed-bid auction with symmetric bidding functions and private value distributions, independent values, payoffs that are increasing with value, and risk neutrality. Therefore, we can apply well-known results from the literature that guarantee the uniqueness of the symmetric equilibrium of Theorem 1 when $N_a \geq 2$, e.g., \cite[Theorem 3]{Maskin84} and \cite{Maskin03}. Furthermore, ties between bidders will almost surely not occur, since different users will experience different channels and therefore different private values with probability 1. Therefore, this equilibrium outcome is also efficient or \emph{Pareto optimal}, since from (\ref{EQ:payoff}) it is easy to see that the bidder with the highest private value always places the highest bid and accordingly wins the auction \cite{Krishna}. $\blacksquare$

We now calculate the revenue generated by the potential helper, keeping in mind that bids as well as payments are non-zero only for weak users with positive private values. As shown in Section~\ref{sec:Extend}, the revenue obtained from ACOPS is the key parameter that stimulates cooperation in the network.

\emph{Theorem 2}: The revenue ${\text{Rev}}\left( {\Omega ^s } \right)$ accrued to the potential helper from the single-object auction game $\Omega ^s$ with rival uncertainty and $N_a$ actual bidders under \emph{both} first-price and second-price rules is given by
\begin{eqnarray}
{\text{Rev}}\left( {\Omega ^s } \right)&=&\sum\limits_{n = 1}^N \left\{ \left( { - 1} \right)^{n - 2}
\left( {\frac{{\Gamma _{1,PH} }}\lambda_1} \right)^{2n}\frac{{\left( {\Gamma _{2,PH} } \right)^n }}
{{\left( {\lambda_2 } \right)^{n - 1} }} \right. \nonumber\\
&&\times\:  \left. {  \left( {\sum\limits_{k = 0}^{n - 2} {C_{(k)}^{(n - 2)} } \frac{1}
{{\left( {n - k - 1} \right)^2 }}} \right)} \right\},
\label{EQ:Theorem2}
\end{eqnarray}
where $C_{(x)}^{(y)}$ is the binomial coefficient.
%\begin{equation}
%{\text{Rev}}\left( {\Omega ^s } \right) = \sum\limits_{n = 1}^N {\left\{ {\left( { - 1} \right)^{n - 2}
%\left( {\frac{{\Gamma _{1,PH} }}
%{{\Gamma _{1,PH}  + \Gamma _{1,BS} }}} \right)^{2n} \frac{{\left( {\Gamma _{2,PH} } \right)^n }}
%{{\left( {\Gamma _{2,PH}  + \Gamma _{2,BS} } \right)^{n - 1} }}} \right.} \left. {  \left( {\sum\limits_{k = 0}^{n - 2} {C_{(k)}^{(n - 2)} } \frac{1}
%{{\left( {n - k - 1} \right)^2 }}} \right)} \right\},
%\label{EQ:Theorem2}
%\end{equation}
%{\text{Rev}}\left( {\Omega ^s } \right) &=&

\emph{Proof}: The proof is shown in the appendix.
\subsection{Outage Probability}
\emph{Theorem 3}: The single-partner ACOPS outage probability for weak user $1$ with target rate $D_1$ is
%\begin{equation}
%P_{out}^s \left( {D_1}  \right)\geq  \left[ {1 - \exp \left( { - \frac{{2^{D_1 }  - 1}}
%{{\Gamma _{1,BS} }}} \right)} \right]\left( {1 - \Psi_1 } \right) + \left[ {1 - \exp \left( { - \frac{{2^{D_1 /2}  - 1}}
%{{\Gamma _{1,PH} }}} \right)} \right] \Psi_1,
% \label{EQ:SingleOutage}
%\end{equation}
\begin{eqnarray}
P_{out}^s \left( {D_1}  \right)&\geq&  \left[ {1 - \exp \left( { - \frac{{2^{D_1 }  - 1}}
{{\Gamma _{1,BS} }}} \right)} \right]\left( {1 - \Psi_1 } \right)\nonumber\\ &&{+}\: \left[ {1 - \exp \left( { - \frac{{2^{D_1 /2}  - 1}}
{{\Gamma _{1,PH} }}} \right)} \right] \Psi_1,
 \label{EQ:SingleOutage}
\end{eqnarray}
where $\Psi _1  \triangleq \sum\limits_{n = 1}^N {\frac{1}
{n}\left( {\frac{{\Gamma _{1,PH} }}
{{\Gamma _{1,PH}  + \Gamma _{1,BS} }}} \right)^{2n} }$.

\emph{Proof}: Assume weak user 1 places a bid and seeks a total rate $R_{1}$ as (\ref{EQ:Cooprate}). As before, define $U^w_1$ as the event that user 1 wins the auction, and $U^l_1$ as the complementary event that user 1 loses the auction. Therefore, the expected outage probability for user 1 with a target rate $D_1$ is
\begin{eqnarray}
  P_{out}^s \left( {D_1 } \right) &=& \Pr \left( {\left[ {C\left( {\gamma _{1,PH} } \right) + C\left( {\gamma _{1,BS} } \right)} \right] < D_1 } \right) \cdot \Pr \left( U^w_1 \right) \hfill \nonumber\\
  &&{+}\: \Pr \left( {C\left( {\gamma _{1,BS} } \right) < D_1 } \right) \cdot \Pr \left( \left(U^w_1\right)^c \right)\\
%\begin{equation}
 &\geqslant& \Pr \left( {2C\left( {\gamma _{1,PH} } \right) < D_1 } \right) \cdot \Pr \left( U^w_1 \right)\nonumber\\
 &&{+}\: \Pr \left( {C\left( {\gamma _{1,BS} } \right) < D_1 } \right) \cdot \Pr \left( U^1_l \right). \label{EQ:Theo3step1}
\end{eqnarray}
User 1's winning probability is
\begin{eqnarray}
\Pr \left( U^w_1 \right) &=& \Pr \left( {U^w_1\mid X_1 > 0} \right) \cdot \Pr \left( {X_1  > 0} \right)\nonumber\\
&&{+}\: \Pr \left( {U^w_1\mid X_1  < 0} \right) \cdot \Pr \left( {X_1  < 0} \right),
\end{eqnarray}
where $\Pr \left( U^w_1\mid X_1  > 0 \right)$ is equivalent to (\ref{EQ:Win}). Since a weak user places a non-zero bid only for positive values, $\Pr \left( U^w_1\mid X_1  < 0 \right)\approx 0$. Therefore, from (\ref{EQ:Prob.Na}) and (\ref{EQ:Win}),
\begin{equation}
\Pr \left( U^w_1 \right) = \sum\limits_{n = 1}^N {\Pr \left( {N_a  = n} \right)} \Pr \left( {U^w_1 \mid N_a  = n} \right)\Pr \left( {X_1  > 0} \right)
\end{equation}
\begin{equation}
\hspace{-0.84 in} = \sum\limits_{n = 1}^N {\frac{1}
{n}\left( {\frac{{\Gamma _{1,PH} }}
{{\Gamma _{1,PH}  + \Gamma _{1,BS} }}} \right)^{2n} }. \label{EQ:Userwinprob}
\end{equation}
Applying (\ref{EQ:NonCoopOutage}) and (\ref{EQ:Userwinprob}) to (\ref{EQ:Theo3step1}) leads to (\ref{EQ:SingleOutage}). $\blacksquare$

\section{MULTIPLE-PARTNER SELECTION}\label{sec:MultiPartner}
We now examine multiple partner-selection by a potential helper capable of assisting more than one weak user at the same time. Applications of such an auction include a cooperative OFDM system \cite{Tarokh}--\cite{LiuOFDM08}, where the potential helper auctions different sub-carriers to different weak users, and a coded cooperation system with multiple simultaneous partners each with a certain block of relayed redundant code-symbols. The rest of this section examines multiple-partner selection in a cooperative OFDM context. The conditions of private values and rival uncertainty outlined in Section III, and best-reply strategies of Proposition 1 are assumed to hold for the multiple-object auction as well. Let the number of subcarriers offered for cooperation by the potential helper be $\tilde K$ whereas user $i$ desires $K_i$ subcarriers to avoid outage, and define $\beta_i={{K_i } \mathord{\left/ {\vphantom {{K_i } {\tilde K}}} \right.
 \kern-\nulldelimiterspace} {\tilde K}}$. Utilizing the concept of marginal opportunity cost from Section IV, the value of subcarrier \emph{k} for weak user \emph{i} is defined as $X_{i,PH}^k  = \gamma _{i,PH}^k/\beta_k  - \gamma _{i,BS}^k $, i.e., the (weighted) difference in SNR between subcarrier \emph{k} on the helper-user channel and subcarrier \emph{k} on the direct-to-BS channel. However, for simplicity we assume $\beta_i=1 \hspace{0.1in} \forall i$ in the remainder of this work.

In a typical system, the number of shared subcarriers is likely to be much greater than the number of partners, e.g., a potential helper with $\tilde K = 128$ subcarriers available for cooperation and five weak users seeking the use of these subcarriers.
\subsection{Multiple-Object Auction Formats for Multiple Subcarriers}
We now list the different multiple-object auction formats available to the potential helper:
\begin{itemize}
\item A series of separate single-object auctions held for each individual subcarrier in turn, resulting in $\tilde K$ separate auctions.
\item A \emph{naive} multiple-object auction, where all subcarriers are sold as separate objects in a single auction.
\item A \emph{pure} bundle auction, where all subcarriers are grouped together into a single package.
\item A \emph{mixed} bundle auction, where subcarriers are grouped into multiple packages of varying size.
\end{itemize}

As an example, consider the simplest format: the naive multiple-object auction, which is a special case of the mixed bundle auction with one subcarrier per bundle. Under the naive multiple-object auction, each weak user obtains the helper-user subcarrier SNR values $\left[ {\gamma _{i,PH}^1 , \ldots ,\gamma _{i,PH}^{\tilde K} } \right]$ and then submits a bid vector $\left[ {b_i^1 , \ldots ,b_i^{\tilde K} } \right]$ for all $\tilde K$ subcarriers. After the $\tilde K$ winning bids out of $\tilde {K}N_a$ total bids have been determined, the payment made for each subcarrier is the second-highest bid placed on that particular subcarrier. The expected payoff for each bidder is $\Pi _{i}  = \sum\nolimits_{j = 1}^{\tilde K} {p_i \left( j \right)} \left[ {x_i^{\left( j \right)}  - \mathop {\max }\limits_{k \ne i} b_k^{\left( j \right)} } \right]$ , where $p_i \left( j \right)$ represents the probability of winning the $j^{th}$ subcarrier. It is apparent that the naive auction format has a prohibitive feedback overhead as the number of subcarriers or bidders increases, therefore we turn to bundled auctions in the sequel.

We first examine the revenue superiority of bundled auctions over the separate and naive auction formats.

\emph{Proposition 3}: For large inter-user SNRs, the distribution of the private value of a bundle is approximated by the density function
\begin{equation}
f^{\left( n \right)} \left( y \right) = \left( {\frac{1}
{{\Gamma _{i,PH} }}} \right)^n \frac{{y^{n - 1} }}
{{\left( {n - 1} \right)!}}e^{ - \frac{1}
{{\Gamma _{i,PH} }}y}.
\label{EQ:Bundlepdf}
\end{equation}

\emph{Proof}: Denote random variable $Y^{\left( n \right)}$ as the sum of $n$ private values, i.e., $Y^{\left( n \right)}  = \sum\nolimits_{j = 1}^n { {X_j } }$. We first obtain the pdf $f^{\left( n \right)} \left( y \right)$ of the sum of private values for a bundle of size $n$. If the difference between helper-user and direct-to-BS SNR for any user is large enough (e.g., greater than 10 dB), the private value pdf can be approximated by the one-sided density for $x_i > 0$ in (\ref{EQ:singlepdf}). Consequently, $f^{\left( n \right)} \left( y \right)$ is the \emph{n}-fold convolution of the individual private value density functions. Denoting the Laplace transform as ${\mathcal{L}} \left\{  \cdot  \right\}$, we have
\begin{equation}
{\mathcal{L}}\left\{ {f^{\left( n \right)} \left( y \right)} \right\}
% = \left[ {{\mathcal{L}}\left\{ {f_x \left( x \right)} \right\}} \right]^n
 = \left[ {\frac{1}{{\Gamma _{i,PH}  + \Gamma _{i,BS} }} \cdot \frac{1}
{{s + \left( {{{ 1} \mathord{\left/
 {\vphantom {{ - 1} {\Gamma _{i,PH} }}} \right.
 \kern-\nulldelimiterspace} {\Gamma _{i,PH} }}} \right)}}} \right]^n,
\end{equation}
where \emph{s} is the complex argument of the Laplace transform. Applying the inverse Laplace transformation \cite[Eqn. 17.13.9]{Ryzhik} and normalizing to ensure unit integral, we obtain (\ref{EQ:Bundlepdf}). $\blacksquare$

Next, we conjecture that a revenue-maximizing potential helper prefers a pure bundle auction for allocation of all subcarriers to separate single-object subcarrier auctions or the naive multiple-object auction, as long as the number of bidders $N_a $ is less than a threshold $N_a^*$.
For exactly two bidders, an elegant proof is given in \cite{Palfrey}. For the case of more than two buyers, the second-price pure-bundle auction is revenue-superior to any efficient auction, if the expectation of the second highest sum of valuations is larger than the sum of the expectations of the second highest valuations, i.e.,
 \begin{equation}
E\left[ {2^{{\text{nd}}} {\text{ highest sum of values}}} \right] > \sum\limits_i^{N_a } {E\left[ {2^{{\text{nd}}} {\text{ highest user }\text{} i\text{} \text{ value}}} \right]}.  \label{EQ:Theorem4}
\end{equation}

%with ${1 \mathord{\left/ {\vphantom {1 {\Gamma _{i,PH} }}} \right. \kern-\nulldelimiterspace} {\Gamma _{i,PH} }} < {\text{Real}}\left( s \right) < {1 \mathord{\left/ {\vphantom {1 {\Gamma _{i,BS} }}} \right. \kern-\nulldelimiterspace} {\Gamma _{i,BS} }}$ as the region of convergence.
The pdf of the second-highest sum of values can be obtained from (\ref{EQ:Bundlepdf}) by using the property of order statistics \cite{Krishna}. Cumbersome integrals obtained for both sides of (\ref{EQ:Theorem4}) were numerically evaluated, and we observe that for the given bundle value distribution, $N_a^*= 7$.
\subsection{Mixed-Bundle Auction}
Having established that a revenue-maximizing potential helper strictly prefers a bundle multiple-object auction up to a moderate number of bidders, we now examine how to adapt the multiple-object auction to our cooperative OFDM network. To simplify the process, we propose a mixed bundle auction format in which the set of all subcarriers is divided into subsets or bundles. Each bundle is now treated as a single sale object. Based upon the results of \cite[Theorem 2]{Jehiel}, we assume that the subcarrier mixed-bundle auction offers a revenue at least as great as the pure-bundle auction. More importantly, the mixed bundle auction allows multiple simultaneous cooperative partners since all subcarriers are not necessarily allotted to a single bidder. The bundle auction contention and allocation process in a cooperative OFDM context is shown in Fig. 2(b).

In the mixed-bundle auction, the surplus bandwidth of the potential helper is divided into \emph{K} bundles, which are simultaneously awarded to $r \leqslant N_a$ bidders offering the \emph{K} highest bids by using a multiple-object auction as follows: Let the $\tilde K$ individual subcarriers be grouped into \emph{K} bundles $H^{\left( 1 \right)} , \ldots ,H^{\left( K \right)}$, with the cardinality of each bundle being $c_k  = \left| {H^{\left( k \right)} } \right|$. The value of the bundle is assumed to be additive, i.e., the value of the bundle to a particular bidder is the sum of the values of the constituent subcarriers. The private-value vector for user $i$ is now ${\mathbf{Y}}_i^B  = \left[ {Y_i^1 , \ldots ,Y_i^K } \right]$, where the superscript $^B$ denotes the bundle auction, and the value of bundle $k$ is $Y_i^k  = \sum\nolimits_{j = i}^{c_k } {X_i^{\left( j \right)} }$. Therefore, in the bundle auction, a group of subcarriers are allocated together to the winning bidder(s).

 The optimal distribution of subcarriers within bundles that maximizes revenue requires a combinatorial search by the helper, which implies prohibitive complexity as $K$ increases. Suboptimal bundling algorithms that guarantee a revenue at least half that of the optimal partitioning are available in the literature \cite{Ghosh07}. However, to reduce complexity as well as the amount of signaling overhead even further, we assume the helper sets the size of each bundle to $\tilde K/N_a$, which is shown to provide a revenue close to that of \cite{Ghosh07}.

\emph{Theorem 4}: The revenue accrued to the potential helper under a mixed-bundle multiple-subcarrier auction game with rival uncertainty is
\begin{equation}
{\text{Rev}}\left( {\Omega ^B } \right) = \sum\limits_{k = 1}^K {\sum\limits_{n = 1}^N {\frac{1}
{N}}  \cdot E_{Y_1^k } \left[ {G_{Y_2^k }^{n - 1} \left( {Y_1^k } \right)} \right] \cdot E_{Y_2^k } \left[ {Y_2^k G_{Y_3^k }^{n - 2} \left( {Y_2^k } \right)} \right]},
 \label{EQ:Theorem5}
\end{equation}
where $G_{Y_i^k}(y)$ is the cumulative distribution function of the $k^{th}$ bundle value $Y_i^k$ equal to
\begin{eqnarray}
G_{Y_i^k } \left( y \right) &=& \frac{1}
{{\left( {\Gamma _{i,PH} } \right)^{c_k } }}\left[ \frac{1}
{{\left( {{1 \mathord{\left/
 {\vphantom {1 {\Gamma _{i,PH} }}} \right.
 \kern-\nulldelimiterspace} {\Gamma _{i,PH} }}} \right)^{c_k  - 1} }} - \left( {e^{\frac{1}
{{\Gamma _{i,PH} }}y} } \right) \right.\nonumber\\
&& {\times}\: \sum\limits_{m = 0}^{c_k  - 1} \left.{\frac{1}
{{m!}}\frac{{y^m }}
{{\left( {{1 \mathord{\left/
 {\vphantom {1 {\Gamma _{i,PH} }}} \right.
 \kern-\nulldelimiterspace} {\Gamma _{i,PH} }}} \right)^{c_k  - m} }}}  \right].
\end{eqnarray}

\emph{Proof}: The cdf $G_{Y_i^k}(y)$ of the bundle value $Y_i^k$ is obtained by integrating (\ref{EQ:Bundlepdf}) using \cite[Eqn. 3.351.1]{Ryzhik}. Each bundle can be treated as a single-object with accrued revenue as derived in (\ref{EQ:Theo2_2ndlast}) of Theorem 2 and assuming ${\Pr \left( {N_a  = n} \right)} = \frac{1}{N}$. Summing over all \emph{K} bundles provides (\ref{EQ:Theorem5}). $\blacksquare$

\emph{Theorem 5}: The outage probability for weak user $i$ in the bundled multiple-subcarrier auction under rival uncertainty is
\begin{equation}
\begin{gathered}
  P_{out}^B \left( {D_1 } \right) \approx \frac{1}
{2}\left( {1 + erf\left( {\frac{{D_1  - \mu _c \left( {\tilde K} \right)}}
{{\sqrt 2 \sigma _c \left( {\tilde K} \right)}}} \right)} \right) \cdot \left( {1 - \Theta _i \left( 1 \right)} \right) \hfill \\
\hspace{0.75 in}   + \frac{1}{2}\left( {1 + erf\left( {\frac{{D_1  - \mu _c \left( {\tilde K + c_1 } \right)}}
{{\sqrt 2 \sigma _c \left( {\tilde K + c_1 } \right)}}} \right)} \right) \cdot \left( {\Theta _i \left( 1 \right)} \right), \hfill \\
\end{gathered} \label{EQ:Theorem6}
\end{equation}
where $\Theta _i \left( 1 \right) \triangleq \frac{1}{N}\sum\limits_{n = 1}^N {E_{Y_1^1 } \left[ {G_{Y_2^1 }^{n - 1} \left( {Y_1^1 } \right)} \right]}$, $\mu_c,\sigma_c$ are defined in (\ref{eq:meanvariance}), and $\operatorname{erf} \left( x \right)$ is the error function.

\emph{Proof}: Define bundle winning probability $\Theta _i \left( K \right) = \Pr \left( {{\text{user }}i{\text{ wins }}K{\text{ bundles}}} \right)$. For the OFDM capacity of (\ref{EQ:OFDMCapacity}), the multiple-partner ACOPS outage probability can be expressed as
\begin{eqnarray}
  P_{out}^B \left( {D_i } \right) &=& \Pr \left( {\tilde C_{i,BS} \left( {\tilde K} \right) < D_i } \right)\cdot\Theta _i \left( 0 \right)\nonumber\\
  &&{+}\: \Pr \left( {\left[ {\tilde C_{i,BS} \left( {\tilde K} \right) + \tilde C_{i,PH} \left( {c_1 } \right)} \right] < D_i } \right)\cdot\Theta _i \left( 1 \right) \nonumber\\
 &&{\ldots}\:  + \Pr \left( {\left[ {\tilde C_{i,BS} \left( {\tilde K} \right) + \tilde C_{i,PH} \left( \sum\nolimits_{i = 1}^K {c_i }
 \right)} \right] < D_i } \right)\nonumber\\
 &&{}\times\:\Theta _i \left( K \right).
\end{eqnarray}
Neglecting the probability of user $i$ winning multiple bundles, we have
\begin{eqnarray}
P_{out}^B \left( {D_i } \right) &\leq& \Pr \left( {\tilde C_{i,BS} \left( {\tilde K} \right) < D_i } \right)\left( {1 - \Theta _i \left( 1 \right)} \right)\nonumber\\
 &&{+} \Pr \left( {\left[ {\tilde C_{i,BS} \left( {\tilde K} \right) + \tilde C_{i,PH} \left( {c_1 } \right)} \right] < D_i } \right)\Theta _i \left( 1 \right). \label{EQ:Theo6step2}
\end{eqnarray}
Following (\ref{EQ:Win}), the probability of user 1 winning bundle $Y_i^k$ is
\begin{eqnarray}
\Theta _1 \left( 1 \right) &=& \sum\limits_{n = 1}^N {\Pr \left( {N_a  = n} \right)}  \cdot \Theta _1 \left( {1|N_a  = n} \right)\\
&=& \frac{1}{N}\sum\limits_{n = 1}^N {E_{Y_1^1 } \left[ {G_{Y_2^1 }^{n - 1} \left( {Y_1^1 } \right)} \right]}.
\end{eqnarray}
 For a finite $J$, the instantaneous OFDM sum capacity $\tilde C_{i,j} \left( {J}\right)$ can be approximated by a normal random variable with mean $\mu _c \left( J \right)$ and variance $\sigma _c^2\left( J \right) $ \cite{Taylor}:
\begin{eqnarray}
\label{eq:meanvariance}
\mu _c\left( J \right)  &=&  - \frac{J}
{{\ln 2}}e^{\left( {{\raise0.7ex\hbox{$1$} \!\mathord{\left/
 {\vphantom {1 {\gamma _0 }}}\right.\kern-\nulldelimiterspace}
\!\lower0.7ex\hbox{${\gamma _0 }$}}} \right)} \operatorname{Ei} \left( { - \frac{1}
{{\gamma _0 }}} \right),\nonumber\\
\sigma _c^2\left( J \right)  &=& J\left[ {\operatorname{var} \left( {C_1 } \right) + 2\sum\limits_{k = 1}^J {\operatorname{cov} \left( {C_1 ,C_k } \right)} } \right],
\end{eqnarray}
where $\operatorname{Ei} \left( x \right) = \int\limits_{ - \infty }^x {\frac{{e^t }}{t}} dt$ is the exponential integral, $\operatorname{var} \left( {C_1 } \right)$ is the variance, and ${\operatorname{cov} \left( {C_k ,C_{k+1} } \right)}$ is the covariance of the instantaneous subchannel capacity random variables.
From the cdf of the normal distribution, we have $\Pr \left( {\tilde C_{i,BS} \left( {\tilde K} \right) < D_i } \right) = \frac{1}
{2}\left( {1 + erf\left( {\frac{{D_i  - \mu _c \left( {\tilde K} \right)}}
{{\sqrt 2 \sigma _c \left( {\tilde K} \right)}}} \right)} \right)$, where $\operatorname{erf} \left( x \right)$ is the error function. Substituting into (\ref{EQ:Theo6step2}), we obtain (\ref{EQ:Theorem6}). $\blacksquare$

We conclude this section with some remarks on the savings in communication overhead afforded by ACOPS.  A fully centralized partner-allocation mechanism requires the availability of complete CSI for all users at the central controller (BS or network access point), which is prohibitive to implement in bandwidth-limited systems. In fact, in networks with asymmetric user channels, reliable CSI reporting to the central controller may not be possible at all for users in a deep fade. The amount of feedback of channel state information (CSI) required for ACOPS will be considerably lower, since users are assumed to be unaware of the CSI of all other users and the partner selection is localized in nature. For the single-object auction process, the amount of unquantized signaling overhead per user (combining rate requests and bids sent to the potential helper) in bits is calculated to be $\left( {\log _2 q_1  + \log _2 b_1 } \right)$, where $b_1 ,{\text{ }}q_1$ are the feedback transmitted by user 1 and are considered to be the same for all users. The global feedback required for the single partner, naive multiple-object auction, bundled multiple-object auction, and centralized multiple-partner selection are listed in Table~II.

\begin{table}[htbp]
\centering
\label{table:Feedback}
\caption{Global Signaling Overhead (bits)}
\begin{tabular}{|l|l|}
\hline
Single partner, ACOPS &    $N\log _2 q_1  + N_a \log _2 b_1$ \\
\hline
Single partner, centralized & $N\log _2 \gamma _{1,BS}  + N!\log _2 \gamma_{1,j}$\\
%%Single partner, centralized & $N\log _2 \gamma _{1,BS}  + N!\log _2 \gamma_{1,j}, j \in \left[ {2,N} \right]$\\
\hline
Multiple partner (naive), ACOPS &	$N\log _2 q_1  + \log _2 N_a b_1^k \tilde K$\\
\hline
Multiple partner (bundled), ACOPS &	$N\log _2 q_1  + \log _2 N_a b_1^k K$\\
\hline
Multiple partner, centralized &	$\tilde K\left(N\log _2 \gamma _{1,BS}  + N!\log _2 \gamma_{1,j}\right)$\\
\hline
\end{tabular}
\end{table}

\section{DISCUSSION}\label{sec:Extend}
The preceding sections have thus far considered one-hop relaying in single stage auctions where the winning bidders are assumed to have sufficient funds to pay the required price. The extension of ACOPS to multi-hop relaying can be achieved using auctions with a resale market, where buyers of goods are free to resell them in a subsequent auction within the same cooperation interval. Next, we consider the more involved scenarios of multiple-stage (sequential) ACOPS, and auctions with more than one potential helper.

\subsection{Multiple Stage ACOPS}
The analysis of repeated auctions over successive time periods necessitates introduction of sequential auction games with budget constraints. First and second-price single-stage auctions with budget constraints are well studied in the literature \cite{Krishna}, however the revenue equivalence principle no longer holds. Sequential single-object auctions with budget constraints is an evolving field of auction theory. Equilibrium strategies have been derived for certain specialized cases such as finite-horizon sequential auctions with an identical good sold in every stage by a fixed seller to budget-constrained bidders without rival uncertainty. Furthermore, most previous studies assume that bidders have utility for only a single unit, i.e., they drop out after winning a particular stage. The conventional modeling of budget constraints also merits mention: budget constraints are either assumed to be a pre-specified constant, or a random variable with a realization revealed to the bidder at every stage \cite{Krishna}.

In the context of partner selection in dynamic wireless networks, our general auction scenario differs in virtually every aspect: non-identical sale objects in successive stages due to changing channel conditions, private values and budgets, infinite time horizon, dynamic entry/exit of users, and absence of a fixed seller.
It is tempting to consider simplifying assumptions in ACOPS in order to make the problem more tractable. For example, assuming budgets are publicly known as in \cite{Honig08} makes it easier for bidders to plan future strategies based on the current bids of competitors. However, the fundamentally unique aspect of our ACOPS game is the fact that users can act as either a bidder or a seller in a given stage, and their budget evolves over time corresponding to their actions and the outcomes of past stages. To our best knowledge, such a sequential auction scenario has yet to be studied by the auction theory community.

Let us assume that in the sequential version of ACOPS, each user attempts to minimize the number of stages in which it suffers from outage over the duration of its presence in the network. The channel conditions and therefore private values of each user are revealed to them in a causal manner at the beginning of every stage. However, the residual budget from the conclusion of a stage carries over to the next one. Therefore, at stage $k$, user $i$ possesses private information $\left(x_i(k),w_i(k)\right)$, where $w_i(k)$ is the available budget.

We first establish that it is always beneficial for a strong user to act as a potential helper. Consider two strong users $A$ and $B$ that enter the game at stage $k$ with an identical initial budget constraint and face $N$ weak users. $A$ is willing to act as a potential helper at stage $k$, whereas $B$ abstains from cooperating with weak users. Since both users are `strong', they possess spare resources by definition, therefore being a helper does not affect $A$'s QoS at the very least. Let us assume that from stage $k+1$ onwards, $A$ and $B$ find themselves in the category of weak users and both begin to bid for cooperation. Now, serving as a helper in stage $k$ enables $A$ to participate in $\eta$ \emph{additional} forthcoming stages after the point at which $B$ exhausts its budget, where $\eta$ can be computed as the ratio of $A$'s expected revenue in stage $k$ to $A$'s expected payments in stages $k+1,k+2,\ldots$ using Theorem 2.

We assume that a reasonable strategy for a user in sequential ACOPS is as follows: if `strong', then act as a helper; if `weak', then bid the minimum of his private value and budget in each stage, which is optimal for the single-stage auction game. A rigorous proof of the optimality of this strategy is beyond the scope of this paper. Alternatively, the bidders may adopt a linear bidding function of the form $c_i(k)x_i(k)$ as proposed in \cite{JunSun}, where the constant $c_i(k)$ is chosen to satisfy the budget constraint $w_i(k)$ at that stage (i.e., the entire budget is placed as a bid regardless of the private value). Intuitively, an aggressive bidding strategy of this kind maximizes the probability of winning a particular stage but risks a negative payoff in that stage, and also leads to a faster depletion of the budget. We compare these two strategies by means of simulation in the next section.

A multi-stage multiple-object auction game with budget constraints is again an open problem in auction theory research, as is the existence and closed-form expressions for equilibria in multiple-object auctions with non-identical goods.
\subsection{Multiple Helpers}
The situation where one weak user can choose between two or more potential helpers in a stage can be studied by using simultaneous auction models \cite{Candale}. In the absence of budget constraints, it is known that the optimal strategy for a bidder is to place bids with all available sellers. For the general multi-stage scenario with budget constraints, the game evolves into a dynamic market game with competition between the multiple sellers at one level, and competition among the bidders as before. Such games have been investigated in the context of dynamic spectrum allocation in wireless networks \cite{Ji06}, for example.

\section{SIMULATION RESULTS}\label{sec:Simul}
The following representative examples are used for simulation purposes. Users are assumed to be quasi-static with a mobile velocity of 3 km/hr to capture the effect of prolonged shadowing.
We refer to max SNR partner selection as partner selection based upon highest instantaneous helper-user SNR $\gamma _{i,PH}$. The cooperative OFDM system is set to $\tilde K = 128$ subcarriers. For all centralized schemes considered, the BS has complete CSI to and between all users, and assigns partners either on a max-min criterion that tries to maximize the lowest user throughput, or an opportunistic throughput-maximizing criterion \cite{Hunter2}. All ACOPS performance results are those achieved at the unique symmetric equilibrium described in previous sections.

\begin{figure}[htbp]
\centering
\includegraphics[width=\linewidth]{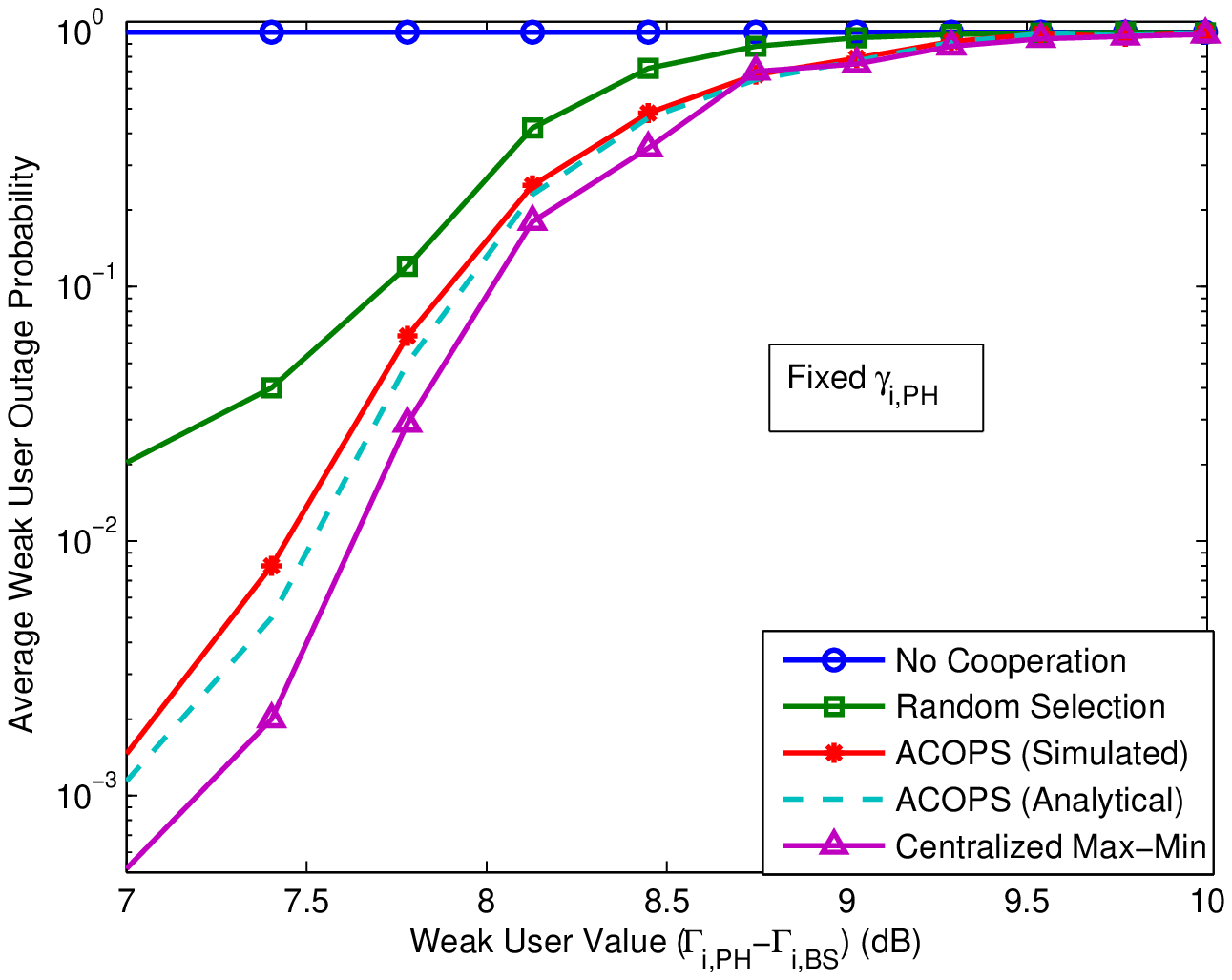}
\caption{Single-object auction game: average weak-user outage probability versus SNR for no cooperation, random selection, maximum instantaneous SNR, auction-theoretic, and centralized-partner selection for $N = 5$ weak users.}
\label{fig:Singlemaxmin}
\end{figure}
 For Fig. \ref{fig:Singlemaxmin}, the average weak user-to-helper SNR is fixed as $\gamma _{i,PH} = 10$ dB, while the direct-link SNR is varied between 7 dB to -20 dB.  We compare the average weak-user outage probability achieved by no cooperation, random selection, simulated and analytical single-object ACOPS, and centralized max-min partner selection with $N = 5$ weak users and target rate $D_i = 10 b/s$. The increasing values of the private value on the abscissa correspond to decreasing values of $\gamma _{i,BS}$. It is observed that in this scenario ACOPS closely mimics the centralized max-min scheduler, which agrees with intuition.

\begin{figure}[htbp]
\centering
\includegraphics[width=\linewidth]{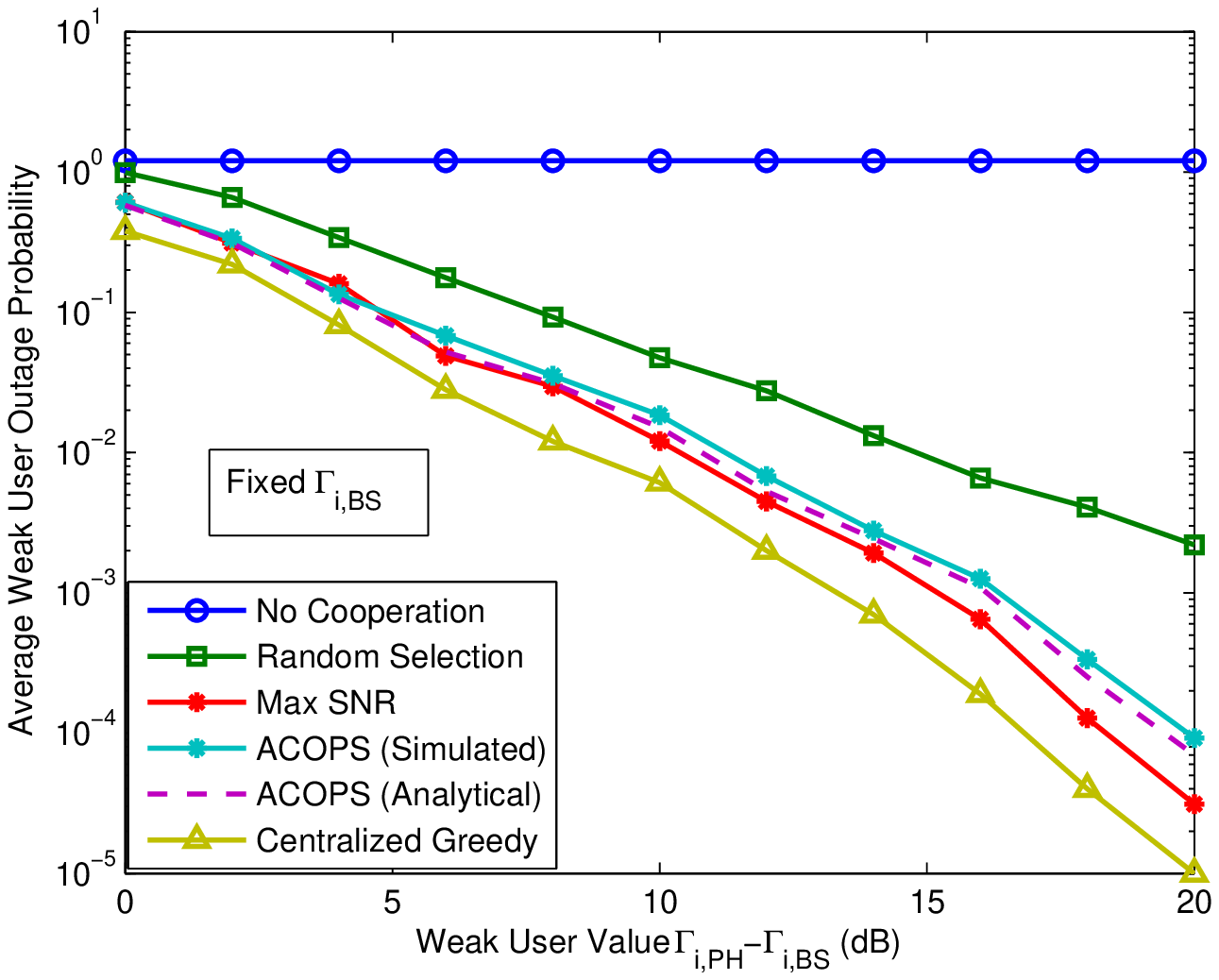}
\caption{Single-object auction game: average weak-user outage probability versus SNR for no cooperation, random selection, maximum instantaneous SNR, auction-theoretic, and centralized-partner selection for $N = 5$ weak users.}
\label{fig:Singlegreed}
\end{figure}
 Next, in Fig. \ref{fig:Singlegreed} the average direct-link SNR is fixed as $\gamma _{i,BS} = 0$ dB, and the weak user-helper SNR is varied between 7 dB to -20 dB. The increasing values of the private value on the abscissa now correspond to increasing values of $\gamma _{i,PH}$.
 It is observed that the ACOPS mechanism performs close to the opportunistic max-SNR selection in the worst direct-to-BS SNR regime due to the choice of the private value expression for $X_i$, while the two curves converge in the high SNR regime. In addition, ACOPS outage probability performance is close to the centralized version even without CSI knowledge of all users.

\begin{figure}[htbp]
\centering
\includegraphics[width=\linewidth]{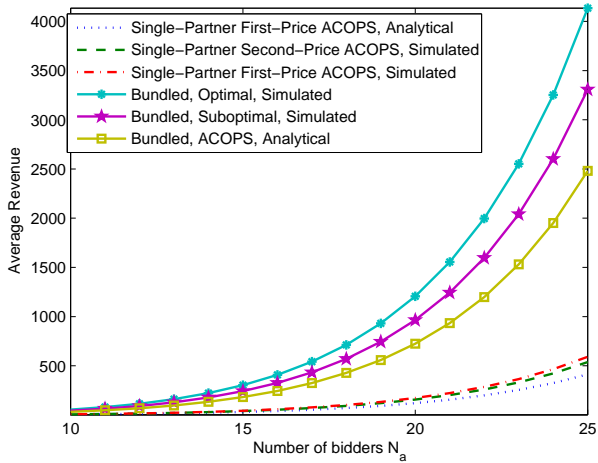}
\caption{Expected revenue accrued by the potential helper when using the single-stage single-partner first-price and second-price and second-price bundled sealed-bid auctions with respect to user-helper SNR. }
\label{fig:Revenue}
\end{figure}
Fig. \ref{fig:Revenue} displays the seller revenue obtained from the first-price and second-price single-object and second-price bundled auctions as the number of bidders increases. The helper-BS link SNR $\gamma _{PH,BS}$ is fixed at 20 dB, and $\gamma _{i,BS} = 0$ dB. For the single-object ACOPS, the simulated and analytical revenue curves are in agreement with theoretical predictions of revenue equivalence. For the bundled ACOPS, the proposed low-complexity bundling partition offers 70\% of the sub-optimal bundling algorithm previously proposed in \cite{Ghosh07}.

\begin{figure}[htbp]
\centering
\includegraphics[width=\linewidth]{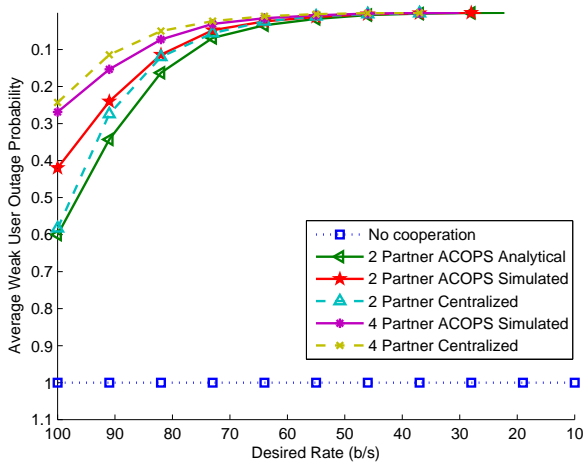}
\caption{Average outage probability for bundle multiple partner selection using ACOPS, centralized-selection, and no cooperation, for number of partners r = 2 and 4,   128 OFDM subcarriers. Total number of weak users $N = 10$.}
\label{fig:Multiple}
\end{figure}
Fig. \ref{fig:Multiple} displays the average weak user outage probability achieved by no cooperation, multiple-object bundle ACOPS and fully centralized multiple-partner selection with respect to \emph{decreasing} desired rates, and $N$ = 10 weak users. The surplus rate $R_c$ of the potential helper is set to be able to accommodate $r = 2$ and 4 simultaneous partners, respectively. Once again, it is observed that the ACOPS outage probability for $r = 2$ and 4 is close to the corresponding centralized outage performance.

\begin{figure}[htbp]
\centering
\includegraphics[width=\linewidth]{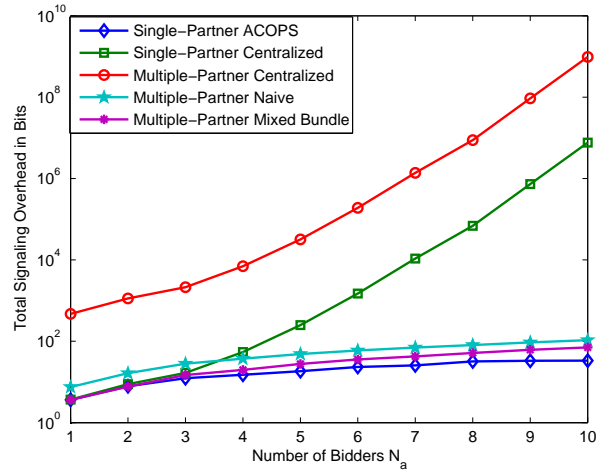}
\caption{Total feedback bits required at potential helper versus number of weak users under single-object ACOPS, naive and bundled multiple-object ACOPS with 512 subcarriers, and feedback bits required at BS under centralized single and multiple-partner selection.}
\label{fig:Feedback}
\end{figure}
Fig.~\ref{fig:Feedback} shows the signaling overhead in bits under single-object and multiple-object (naive and bundle formats) ACOPS, and a BS under centralized single and multiple partner selection versus number of weak users $N$. For the bundle multiple-object auction, the number of bundles $K$ is equated to the number of bidding weak users $N$ for simplicity. The ACOPS mechanism requires significantly reduced signaling overhead in both single and multiple-partner scenarios, e.g., by a factor of 9 for single-partner ACOPS versus centralized selection with $N=5$ weak users.

\begin{figure}[htbp]
\centering
\includegraphics[width=\linewidth]{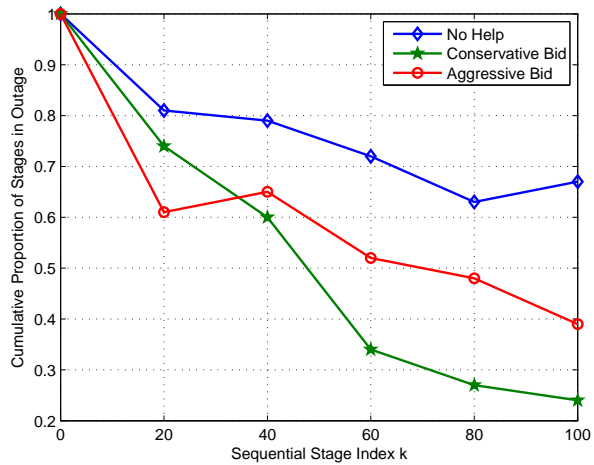}
\caption{Cumulative fraction of stages in outage for multi-stage single-partner ACOPS. $D_i=6$ (bits/s/Hz) $\forall i$, with 6 total users in the network per stage.}
\label{fig:Sequential}
\end{figure}
Finally, Fig.~\ref{fig:Sequential} displays the cumulative fraction of stages in outage for multi-stage single-partner ACOPS over 100 successive stages. The initial budget constraint for all users is set to $w_0=5000$.  The legends ``Conservative Bid" and ``Aggressive Bid" denote the strategies of bidding as $b_i\left(x_i(k),w_i(k)\right)=\min\left(x_i(k),w_i(k)\right)$ versus spending the entire budget per stage as $b_i\left(x_i(k),w_i(k)\right)=w_i(k)$, respectively, while always serving as a potential helper when in a strong state in both cases. The legend ``No Help" corresponds to the case where a user \emph{never} acts as a helper during his strong state, and bids conservatively when in a weak state. The channel realizations are independent across users and stages, while budgets carry over. The required data rate is set such that users are in outage roughly 70\% of the time without any cooperation. We observe that the users that do not assist others exhaust their budget very quickly and thereafter are unable to participate in auctions when weak. The conservative bidding strategy outperforms the aggressive strategy by allowing the users to bid successfully in a greater number of auction stages.

\section{Conclusion}\label{sec:Conclus}
A decentralized and low-complexity cooperative partner-selection scheme based on auction theory was developed and analyzed for dynamic ad hoc networks. Best-response strategies were characterized for the second-price sealed-bid auction game with uncertainty in the number of bidders, followed by the multiple-object bundle auction game. The proposed cooperative communication protocol with minimum network condition information was shown to achieve outage performance close to a centralized partner-selection scheme with complete CSI, even under adverse channel conditions. The auction-theoretic partner-selection strategy was also shown to significantly reduce the signaling overhead involved in configuring cooperative communications, which will translate into improved bandwidth and power efficiency for mobile users. Although the present study featured a cooperative network with a central transceiver as the final destination, the proposed partner-selection scheme can be extended to accommodate ad hoc networks with multiple transmitter-receiver pairs.

\appendix[Proof of Theorem 2]
% or
\emph{Proof}: Consider ACOPS under second-price auction rules. Let $\Pr \left( {N_a  = n} \right)$ be the probability of $n$ actual bidders with positive private values, and consider bidder 1 with bid $b_1 \left( {x_1 } \right) = x_1$, $x_1 > 0$. Given that the $n$ private values are i.i.d with distribution $F_{X_i}$, we have
\begin{equation}
 \Pr \left( {N_a  = n} \right) = \prod\limits_{i = 1}^n {\left( {1 - F_{X_i } \left( 0 \right)} \right)}  = \left( {\frac{{\Gamma _{1,PH} }}
{{\Gamma _{1,PH}  + \Gamma _{1,BS} }}} \right)^n. \label{EQ:Prob.Na}
 \end{equation}
Define $U^w_1$ as the event that bidder 1 wins the auction, and $U^l_1$ as the complementary event that bidder 1 loses the auction. The probability that bidder 1 possesses the highest bid is given by
\begin{equation}
\Pr \left( U^w_1 \right) = \sum\limits_{n = 1}^N {\Pr \left( {N_a  = n} \right)} \Pr \left( {U^w_1\mid N_a  = n} \right). \label{EQ:Win}
\end{equation}
Now, the winning probability for bidder 1 conditioned upon the number of actual bidders is
\begin{equation}
\Pr \left( {U^w_1 \mid N_a  = n} \right) = \Pr \left( {X_2  \leq X_1 , \ldots ,X_n  \leq X_1 \mid N_a  = n} \right)
\end{equation}
%\begin{equation}
% = \int\limits_0^\infty  {\Pr \left( {X_2  \leq x_1 , \ldots ,X_n  \leq x_1 \mid N_a  = n,X_1  = x_1 } \right)f_{X_1} \left( {x_1 } \right)} dx_1
%\end{equation}
\begin{equation}
 = \int\limits_0^\infty  {\left[ {F_{X_2 } \left( {x_1 } \right)} \right]^{n - 1} f_{X_1} \left( {x_1 } \right)} dx_1  = E_{X_1 } \left[ {F_{X_2}^{n - 1} \left( {X_1 } \right)} \right]. \label{EQ:step1}
\end{equation}
The expected revenue for the seller from bidder 1 alone is derived next. Assuming $X_2  = \mathop {\max }\limits_{j,j \ne 1} X_j$, the conditional payment made by winning bidder 1, $E\left[ {{\text{Rev}}\left( {\Omega ^s } \right)\mid N_a  = n} \right]$, can be written as
\begin{equation}
%\begin{gathered}
E_{X_2 } \left[ { {\text{$2^{nd }$ highest bid  }}\mid {\text{bidder 2 is the $2^{nd }$ highest bidder  }}} \right]
\end{equation}
\begin{eqnarray}
&=& E_{X_2 } \left[ {2^{nd } {\text{ highest bid  }}\mid X_{i=3,\ldots,n}  \leq X_2 , X_1  > X_2, N_a  = n} \right]\nonumber\\
 &&{}\cdot \Pr \left( {U^w_1 \mid N_a  = n} \right)\\
%\end{gathered}
%\begin{equation}
% = \int\limits_0^\infty  {x_2} \Pr \left( {X_3  \leq X_2 , \ldots ,X_n  \leq X_2 |X_2  = x_2 ,X_1  > x_2 } \right)f_{X_2} \left( {x_2 } \right)dx_2  \cdot \Pr \left( {{\text{bidder 1 wins}}|N_a  = n} \right)
%\end{equation}
%\begin{equation}
 &=& E_{X_2 } \left[ {X_2 F_{X_3}^{n - 2} \left( {X_2 } \right)} \right]\Pr \left( {U^w_1 \mid N_a  = n} \right). \label{EQ:step2}
\end{eqnarray}
Therefore, using (\ref{EQ:step1}) and (\ref{EQ:step2}), the expected revenue at the potential helper averaged over all actual bidders is
 \begin{eqnarray}
{\text{Rev}}\left( {\Omega ^s } \right) &=& \sum\limits_{n = 1}^N \left\{{\Pr \left( {N_a  = n} \right)}  \cdot n E_{X_1 } \left[ {F_{X_2}^{n - 1} \left( {X_1 } \right)} \right] \right.\nonumber\\
&&{\cdot}\: \left. E_{X_2 } \left[ {X_2 F_{X_3}^{n - 2} \left( {X_2 } \right)} \right] \right\}. \label{EQ:Theo2_2ndlast}
\end{eqnarray}
  Under the i.i.d private value assumption, using $F_{X_2} = F_{X_1}$ and \cite[Eqn. 3.312.1]{Ryzhik}, we obtain $E_{X_1 } \left[ {F_{X_2}^{n - 1} \left( {X_1 } \right)} \right] =A_1^n  \cdot \beta \left( {1,n} \right)$, where $A_i = \frac{{\Gamma _{i,PH} }}{{\Gamma _{i,PH}  + \Gamma _{i,BS} }}$, and $\beta \left( {a,b} \right) = \int_0^1 {t^{a - 1} \left( {1 - t} \right)^{b - 1} dt}$ is the beta function. Since the beta function has integer arguments, it can be simplified to $\beta \left( {1,n} \right) = \frac{{\left( {n - 1} \right)!}}{{n!}} = \frac{1}{n}$. Similarly, $E_{X_2 } \left[ {X_2 F_{X_3}^{n - 2} \left( {X_2 } \right)} \right] = \left( { - 1} \right)^{n - 2} A_2^{n - 1}  \Gamma _{2,PH}  \sum\nolimits_{k = 0}^{n - 2} {C_{(k)}^{(n - 2)} } \frac{1}{{\left( {n - k - 1} \right)^2 }}$, using $F_{X_3} = F_{X_2}$ and \cite[Eqn. 3.432.1]{Ryzhik} with a change of variable. Substituting along with (\ref{EQ:Prob.Na}) into (\ref{EQ:Theo2_2ndlast}), we obtain the result of (\ref{EQ:Theorem2}).

To show that ACOPS under first-price auction rules has the same revenue, we invoke the Revenue Equivalence Theorem \cite{Tirole}, which guarantees that both auction games should provide the same revenue even under rival uncertainty, as long as (i) a bidder with the lowest feasible value has an expected payment of zero, (ii) the bidder with the highest value is allocated the object, and (iii) bidders share a private value distribution with a monotone hazard rate $\lambda \left( x \right) = \frac{{f_X \left( x \right)}}{{1 - F_X \left( x \right)}}$. Conditions (i),(ii) are satisfied by our auction model assumptions in Section III and Proposition 1. The hazard rate of the given private value distribution is unimodal and is monotonically increasing (decreasing) for negative (positive) values. Therefore, the first-price ACOPS has the same revenue as the second-price ACOPS given by (\ref{EQ:Theorem2}). $\blacksquare$

\end{document}